\documentstyle[11pt,epsfig,axodraw]{article}
\def\de{\partial}
\def\oh{\frac{1}{2}}
\begin{document}
\begin{titlepage}
\begin{flushright}
DFTT 58/97\\
September 1997
\end{flushright}
\vskip0.5cm
\begin{center}
{\Large\bf 
Interaction effects in the spectrum\\* 
\vskip0.5cm
of the three--dimensional Ising model
}\\ 
\end{center}
\vskip 0.6cm
\centerline{P. Provero}
\vskip 0.6cm
\centerline{\sl Dipartimento di Fisica
Teorica dell'Universit\`a di Torino}
\centerline{\sl Istituto Nazionale di Fisica Nucleare, Sezione di Torino}
\centerline{\sl via P.Giuria 1, I--10125 Torino, Italy
\footnote{e--mail:provero@to.infn.it}}
\vskip 0.6cm
\begin{abstract}
The two--point correlation functions of statistical models show in general 
both poles and cuts in momentum space. The former correspond to the spectrum of
massive excitations of the model, while the latter originate from interaction
effects, namely creation and
annihilation of virtual pairs of excitations. We discuss the effect of 
such interactions
on the long distance behavior of correlation functions in configuration space,
focusing on certain time--slice operators which are
commonly used to extract the spectrum. For the $3D$ Ising model in
the scaling region of the broken--symmetry phase, a one--loop calculation shows
that the interaction effects on time--slice correlations is non negligible for
distances up to a few times the correlation length, and should therefore be
taken into account when analysing Monte Carlo data.
\end{abstract}
\end{titlepage}
\setcounter{footnote}{0}
\def\thefootnote{\arabic{footnote}}
\section{Introduction}
When studying a statistical model, one is often interested in determining the
spectrum of massive excitations, {\em i.e.} the eigenvalues of the transfer
matrix. For many interesting models this cannot be done exactly, and one has to
rely on approximate methods or numerical Monte Carlo calculations.
\par
The observables which are most suitable to investigate the spectrum of a model
are the two--point correlation functions of operators: their long distance 
behavior is directly related to the spectrum. This is especially evident in
momentum space, where each pole of the correlation function corresponds to a
massive excitation, and therefore to an eigenvalue of the transfer matrix.
\par
In general, however, the momentum space correlators 
will have not only poles but
also cuts, signaling the possibility of creating and annihilating virtual pairs
of excitations. For example if the effective Hamiltonian for the order parameter
$\phi$ includes a $\phi^3$ interaction, the Feynman diagram
\begin{center}\begin{picture}(250,30) (0,0)
\Line(70,15)(90,15)
\Line(120,15)(140,15)
\BCirc(105,15){15}
\end{picture}\end{center}
will generate a cut in the $\langle\phi\phi\rangle$ correlator in momentum
space.
\par
Since, in general, Monte Carlo simulations give direct access to
configuration space correlation functions rather than their Fourier transforms,
it is interesting to study the effect of such interactions 
on the long distance  behavior of 
configuration space correlators.
The purpose of this work is to compute this effect in the scaling region of the
broken symmetry phase of the $3D$ Ising model, 
where actual calculations can be
performed in the framework of renormalized Euclidean quantum field theory.
\par
A strong motivation for this analysis is provided by the continuous improvements
in the accuracy of Monte Carlo simulations: recent advances in both computer
performances and simulation algorithms allow us to obtain numerical data of
unprecedented precision. Their analysis requires more sophisticated
theoretical tools as finer effects become observable. 
We will find that the effects that are the object of this study typically 
account for
about one percent of the correlators in the region of physical interest:
this is actually an order of
magnitude larger than the statistical uncertainties typical of recent Monte
Carlo studies of the $3D$ Ising model.
\par
The paper is organized as follows: in Sect. 2 we introduce time--slice
operators, which are particularly suitable for the Monte Carlo study of the
spectrum of a statistical model. In Sect. 3 we  compute the correlators of
such operators in the Ising model using $\phi^4$ field theory at one loop in
three Euclidean dimensions. In particular the interaction effects can be
evaluated and expressed in terms of exponential integral functions.
In Sect. 4 we comment on the relevance of these  effects to certain
universal amplitude ratios, while
Sect. 5 is devoted to the discussion of the results 
and their implications for the analysis of
Monte Carlo data.
\section{Time--slice correlators}
The spectrum of massive excitations of a statistical model can be obtained by
studying the long distance behavior of two--point correlation functions.
Consider for example the Gaussian model, with Hamiltonian
\begin{equation}
H=\int d^dx\left(\frac{1}{2}\de_\mu\phi\de_\mu\phi+\frac{m^2}{2}\phi^2\right)
\ .
\end{equation}
The two--point correlation function is
\begin{eqnarray}
\langle \phi(x) \phi(y)\rangle&=&\int\frac{d^d p}{(2\pi)^d}\frac
{e^{ip(x-y)}}{p^2+m^2}\nonumber\\
&=&\frac{1}{(2\pi)^{d/2}}\left(\frac{\left|x-y\right|}
{m}\right)^{1-d/2}K_{1-d/2}\left(m\left|x-y\right|\right)
\ ,
\end{eqnarray}
where $K_{1-d/2}$ is a modified Bessel function. Therefore asymptotically for 
$\left|x-y\right|\to\infty$
\begin{equation}
\langle \phi(x) \phi(y)\rangle\sim{\rm const}\ m^{\frac{d-3}{2}}
\left|x-y\right|^{-\frac{d-1}{2}}e^{-m\left|x-y\right|}\ .
\label{corr}\end{equation}
We see that the long distance 
behavior of correlators can
be used to extract the value of the mass $m$, which in this case is obviously
the only state in the spectrum.
\par
In practice, it is more convenient to define time--slice operators
\begin{equation}
S(t)=\frac{1}{L^{d-1}}\int dx_1 \dots dx_{d-1} \phi(x_1,\dots, x_{d-1}, t)
\end{equation}
to obtain a purely exponential behavior of correlations. For example in
the Gaussian model it is easy to see that
\begin{equation}
\langle S(t) S(0) \rangle=\frac{1}{L^{d-1}}e^{-m|t|}
\ .\label{cslice}
\end{equation}
When considering a non--trivial model, it is customary to
generalize
Eq.(\ref{cslice}) to
\begin{equation}
\langle S(t) S(0) \rangle=\sum_k c_k e^{-m_k|t|}
\ .\label{cslice2}
\end{equation}
By fitting the values of the time--slice correlations with Eq.(\ref{cslice2})
one can extract the values of a certain number of low--lying states, depending
on the precision of the available data. 
\par
Each exponential in the r.h.s. of Eq.(\ref{cslice2}) corresponds to a pole 
in the Fourier transform
of $\langle\phi(x)\phi(y)\rangle$. 
However, since a non--trivial model certainly involves interactions, Eq.
(\ref{cslice2}) must be modified to take into account their contribution:
in the next Section we will compute the
time--slice correlator $\langle S(0) S(t) \rangle$ for the $3D$ Ising model in
the broken symmetry phase, where a $\phi^3$ interaction is present in the
effective Hamiltonian so that cuts appear in the two--point function 
already at one loop. The calculation will teach us how to
modify Eq.(\ref{cslice2}) to take into account the effect of
production/annihilation of virtual pairs of excitations. 
\section{The case of the $3D$ Ising model}
It is a widely accepted conjecture that the $3D$ Ising model is in the same
universality class as $\phi^4$ field theory. This allows us to use renormalized
$3D$ quantum field theory to study the Ising model in the scaling region, where
lattice effects become negligible and universality holds. This program was
initiated by Parisi \cite{pa} and has been vigorously pursued to study several
aspects of the Ising model \cite{qf,2lo,3lo}. 
The agreement between field theoretical
calculations and Monte Carlo results is satisfactory.
\par
Therefore, from now on we will consider the $3D$ Euclidean field theory 
defined by the action (effective hamiltonian, in the language of statistical
mechanics):
\begin{equation}
S=\int d^3 x \left[\oh\de_\mu\phi\de_\mu\phi+\frac{g}{24}\left(
\phi^2-v^2\right)^2\right]
\end{equation}
We are interested in the two--point connected function
\begin{equation}
G(x-y)=\langle\phi(x)\phi(y)\rangle-\langle\phi(x)
\rangle\langle\phi(y)\rangle\ . \label{g}
\end{equation}
The perturbative expansion must be performed around one of the stable classical
solution, say $\phi=v$. Defining a fluctuation field $\varphi=\phi-v$ a 
$\varphi^3$ term appears in the Lagrangian, with a coupling proportional to
$\sqrt{g}$ (for details about the perturbative expansion in the broken symmetry
phase see {\em e.g.} Ref.\cite{2lo}). 
The correlation function (\ref{g}) is then given at
one loop by the sum of the following Feynman diagrams:
\begin{center}\begin{picture}(250,52) (0,0)
\Text(20,25)[]{G(x-y) =}
\Line(60,25)(90,25)
\Text(75,0)[]{(a)}
\Text(100,25)[]{+}
\Line(110,25)(140,25)
\BCirc(125,33){8}
\Text(125,0)[]{(b)}
\Text(150,25)[]{+}
\Line(160,25)(190,25)
\Line(175,25)(175,34)
\BCirc(175,42){8}
\Text(175,0)[]{(c)}
\Text(200,25)[]{+}
\Line(210,25)(220,25)
\BCirc(230,25){10}
\Line(240,25)(250,25)
\Text(230,0)[]{(d)}
\end{picture}\end{center}
Using dimensional regularization we find
\begin{equation}
G_d(x-y)=\int\frac{d^dp}{(2\pi)^d}\tilde{G}_d(p)e^{ip(x-y)}
\end{equation}
with
\begin{equation}
\tilde{G}_d(p)=\frac{1}{p^2+m^2}+\frac{g}{m^2}\left(\frac{m^2}{4\pi}
\right)^{d/2}\frac{\Gamma\left(1-\frac{d}{2}\right)}
{\left(p^2+m^2\right)^2}
+\frac{3m^2g}{2}\frac{F_d(p)}{\left(p^2+m^2\right)^2}
\label{gtilde}
\end{equation}
and
\begin{equation}
F_d(p)=\frac{\Gamma(2-d/2)}{(4\pi)^{d/2}}
\int_0^1 dx \left[m^2+x(1-x)p^2\right]^{d/2-2}\ .
\end{equation}
\par
The first term in the r.h.s. of Eq.(\ref{gtilde}) is the tree level
contribution with a pole in $p^2=-m^2$. 
The remaining diagrams are interaction effects:
the second term, corresponding to diagrams (b)+(c), produces
shift in the location of the pole, {\em i.e.} a quantum correction to the
physical mass. 
The third term is the interesting one for our purpose: besides
providing another quantum correction to the physical mass, it has a cut in
$p^2=-4m^2$. Indeed, this term corresponds to diagram (d)
in the expansion of  $G(x-y)$ namely 
to the production and annihilation of a virtual pair of particles.
\par
The theory must be renormalized to be compared with experimental or Monte Carlo
results. Notice that analytic continuation to $d=3$ gives a finite expression
for $\tilde{G}(p)$ without need for any subtraction. This is a peculiarity of
dimensional regularization for odd $d$ which disappears when higher loop
effects are taken into account. It implies that the renormalized parameters in
the MS scheme coincide at one loop with the bare parameters. 
\par
However the MS scheme is not particularly suited for direct comparison of
field theoretic calculations with experimental or Monte Carlo 
data. A more convenient scheme,
where the renormalized parameters of the field theory have a direct lattice
interpretation was introduced in Ref.\cite{ren}. 
The renormalized parameters $\phi_R$, $m_R$, $g_R$ in this
scheme were computed at three-loop order in \cite{3lo}, to which we refer the
reader for the definitions and expressions of the renormalized parameters. 
In particular the renormalized mass $m_R^2$ is defined as the momentum space
correlation at $p=0$, and coincides with the inverse second--moment correlation
length.
Here
we just need the expression of the connected two--point function of renormalized
fields:
\begin{equation}
G_R(x-y)=\langle\phi_R(x)\phi_R(y)\rangle_c=\int\frac{d^3p}{(2\pi)^3}
\tilde{G}_R(p)e^{ip(x-y)}
\end{equation} 
with, denoting with $u_R=g_R/m_R$ the dimensionless renormalized coupling,
\begin{eqnarray}
\tilde{G}_R(p)&=&\left(1+\frac{u_R}{64\pi}\right)\left\{\frac{1}
{p^2+m_R^2\left(1-\frac{3u_R}{64\pi}
\right)}-\frac{m_R^2u_R}{4\pi}\frac{1}{(p^2+m_R^2)^2}+\right.\nonumber\\
&&\left.\frac{3 m_R^3 u_R}{8\pi}\frac{1}{(p^2+m_R^2)^2}\frac{1}{\sqrt{p^2}}
\arctan\left(\frac{\sqrt{p^2}}{2m_R}\right)\right\}
\end{eqnarray}
\par
Defining the time--slice operators as
\begin{equation}
S(t)=\frac{1}{L^2}\int dx_1 dx_2 \phi_R(x_1,x_2,t)
\end{equation}
we have 
\begin{equation}
\langle S(t) S(0)\rangle_c=\frac{1}{L^2}\int\frac{dp}{2\pi}e^{ipt}
\tilde{G}_R(0,0,p)
\end{equation}
After a simple calculation we obtain for $t>0$
\begin{eqnarray}
\langle S(t) S(0)\rangle&=&
\frac{1}{2 m_R L^2} e^{-m_{ph} t}
\left[1+\frac{u_R}{128 \pi}\left(24 \log 3-27\right)\right]\nonumber\\
&&\ +\frac{3 u_R}{16 \pi L^2 m_R}\int_{2 m_R}^\infty
d\mu\frac{e^{-\mu t}}
{\mu\left(1-\frac{\mu^2}{m_R^2}\right)^2}\ .\label{zeromom}
\end{eqnarray}
where $m_{ph}$ is the physical mass, defined as the location of the zero of the
inverse correlator in momentum space $G^{-1}(p)$
\begin{equation}
m_{ph}^2=m_R^2\left[1+\frac{u_R}{64\pi}\left(13-12\log 3\right)\right]
\label{mphys}
\end{equation}
The integral appearing in Eq.(\ref{zeromom}) can be expressed in terms of
exponential integral functions
to give 
\footnote{The appearance of an exponential term $e^{-2mt}$ in 
Eq.(\ref{ei}) could be misleading: due to a cancellation between this term and
the ${\rm Ei}$ functions the asymptotic behavior of the sum in square brackets 
is actually $e^{-2mt}/t$.}

\samepage{
\begin{eqnarray}
&&\langle S(t) S(0)\rangle_c=\frac{1}{2 m_R L^2} e^{-m_{ph} t}
\left[1+\frac{u_R}{128 \pi}\left(24 \log 3-27\right)\right]\nonumber
\\
&&\ +\frac{3 u_R}{16 \pi L^2 m_R}\left[\frac{e^{-2m_{ph}t}}{6}
+\frac{m_{ph}t+2}{4}e^{-m_{ph}t}{\rm Ei}(-m_{ph}t)\right.
\nonumber\\
&&\left.\ +\frac{2-m_{ph}t}{4}e^{m_{ph}t} {\rm Ei}(-3 m_{ph}t)
-{\rm Ei}(-2 m_{ph}t)\right]\label{ei}
\end{eqnarray}}

Eq.(\ref{ei}) is our main result: it gives the contribution of interaction
effects to the
correlation function of time--slice operators as a modification of the simple
exponential behavior (\ref{cslice2}). 
\section{The universal amplitude ratio $\xi/\xi_{2nd}$}
Universal amplitude ratios are certain dimensionless combinations of observables
that are predicted to be universal at criticality
(for a comprehensive review see Ref.\cite{priv}).
Among these it is of particular relevance to us the ratio $\xi/\xi_{2nd}$ 
of the second--moment
correlation length and the "true" correlation length ({\em i.e.} the inverse of
the physical mass). 
This ratio defines two universal amplitude ratios, $F_+(F_-)$, when
the critical limit is taken from the symmetric (broken symmetry) phase.
It can be shown \cite{ur} that the presence of higher masses in the 
spectrum implies $F_{\pm}>1$. 
However the corresponding analysis in $\phi^4$ theory
shows that the converse is not true: a value of $F_-$ greater than one 
does not necessarily indicate the presence of higher mass states, but can simply
be a signal of non--trivial interaction effects like the ones studied in this
work. 
\par
In fact,
since the renormalized mass $m_R$ is
defined as $1/\xi_{2nd}$ (see \cite{ren}), from Eq.(\ref{mphys}) we see that 
at one loop in $\phi^4$ theory, \cite{unpu}
\begin{equation}
F_-=1-\frac{u_R^*}{128\pi}(13-12\log 3)=1.00668(3)\ ,
\end{equation}
in good agreement with the Monte Carlo result \cite{ur}
\begin{equation}
F_-=1.009(5)
\end{equation}
Moreover, it is easy to see that the non--trivial contribution to
$F_-$ comes exclusively from  diagram (d) in the expression of $G(x-y)$,
namely the diagram which produces the cut. 
\par 
It would be interesting to study similar effects in the symmetric phase:
however we expect them to appear with the diagram
with the diagram
\begin{center}\begin{picture}(250,30) (0,0)
\Line(70,15)(140,15)
\CArc(105,15)(15,0,180)
\CArc(105,15)(15,180,360)
\end{picture}\end{center}
and therefore only at two--loop level. This provides 
a qualitative explanation for
the fact that in the symmetric phase 
the corresponding amplitude ratio $F_+$ is
known to be much smaller than $F_-$: 
a strong coupling expansion gives \cite{pe}
\begin{equation}
F_+=1.00023(5)
\end{equation}
while Monte Carlo calculations give an upper bound \cite{ur}
\begin{equation}
F_+<1.0006
\end{equation}
\section{Discussion}
The relevance of the effect we have just computed can be best appreciated 
by considering the ratios 
\begin{equation}
R(t)=-\log\frac{\langle S(t+1)S(0)\rangle_c}{\langle S(t)S(0)\rangle_c}
\label{ratio}
\end{equation}
For a purely exponential behavior, $R(t)$ is identically equal to $m_{ph}$,
while $R(t)$ as given by Eq.(\ref{ei}) is plotted in Fig. 1, where we have used
the Monte Carlo estimate \cite{ur} 
\begin{equation}
u_R^*=14.3(1)
\end{equation}
for the value of the dimensionless renormalized coupling in the continuum limit
\footnote{The variation of $u_R$ is anyway very slow in the whole scaling
region, being governed by corrections to scaling.}
and we have set $m_{ph}=1$ ({\em i.e.} we are measuring distances in units of
the correlation length).
\begin{figure}[h]
\begin{center}
\mbox{\epsfig{file=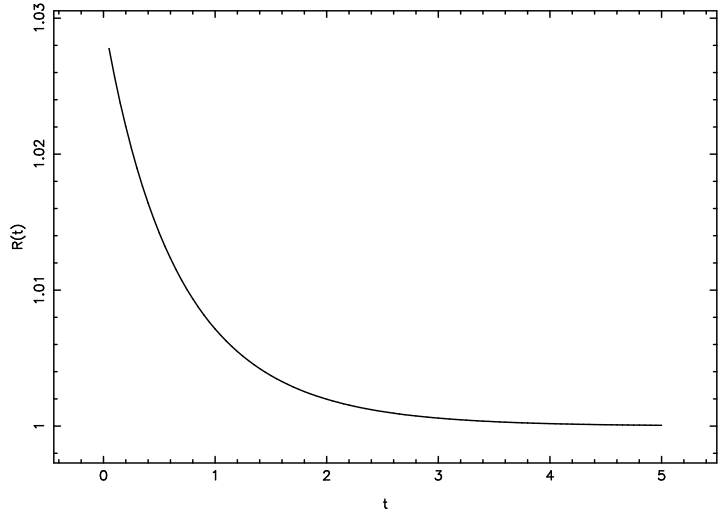,height=7.5cm,width=10.cm}}
\vskip 2mm
\caption
{\it 
The ratio $R(t)$ defined in Eq.(\ref{ratio}) as predicted by Eq.(\ref{ei})
for $m_{ph}=1$.
A purely exponential behavior would give $R(t)=1$ identically.
}
\end{center}
\end{figure}
\par
The figure shows that $R(t)$ is appreciably different from $1$ for distances of
the order of a few times the correlation length. 
For example at $t=1$ the interaction effect is $\sim 0.8\%$ of $R$.
State--of--the--art Monte Carlo simulations give statistical uncertainties about
ten times smaller for the same quantity \cite{ur}. The magnitude of the effect
becomes comparable with the statistical uncertainties at $t\sim 2.5$.
\par
The specific form of Eq.(\ref{ei}) suggests one more reason why the correction
must be taken into account when analysing numerical data. 
In fact, the behavior of the correlators when interaction 
contributions
are included mimics very closely the contribution of a higher mass in the
spectrum. Consider the ratio (\ref{ratio}) when the contribution of two poles is
included and the cut is neglected:
\begin{equation}
\tilde{R}(t)=-\log\frac{e^{-m(t+1)}+\alpha e^{-m^\prime(t+1)}}
{e^{-mt}+\alpha e^{-m^\prime t}}
\end{equation}
We have verified that by adjusting the parameters $\alpha$ and $m^\prime/m$ one
can make $\tilde{R}(t)$ look very similar to the cut contribution:
relative uncertainties in the data 
of less than one part in $10^4$ would be needed to
resolve the difference. The "best fit" value is about $m^\prime/m=2.4$.
Therefore the effect of interaction can easily be mistaken for a higher mass
state with $m^\prime\sim 2.4\  m$.
\par
A more complete investigation of these issues, including a high--precision Monte
Carlo analysis, is currently being pursued and will be published elsewhere 
\cite{ew}.
\vskip2.cm
\noindent
I would like to thank M. Caselle and F. Gliozzi for many helpful discussions.

\end{document}